\documentclass[twocolumn,prd,groupedaddress,nofootinbib,showpacs]
{revtex4}

\usepackage{latexsym} % D'Alambertian operator \Box
\usepackage{amssymb}  % D'Alambertian operator \square
\usepackage{graphicx} % Include figure files
\usepackage{dcolumn}  % Align table columns on decimal point
\usepackage{bm}       % bold math

\begin{document}

\title{On the trace anomaly and the energy-momentum conservation of
quantum fields at D=2 in classical curved backgrounds}

\author{M. Alves}
\email{msalves@if.ufrj.br}
\author{J. Barcelos-Neto}
\email{barcelos@if.ufrj.br}
\affiliation{Instituto de F\'{\i}sica\\
Universidade Federal do Rio de Janeiro, RJ 21945-970 -- Brazil}

\date{\today}

\begin{abstract}
We study the conformal symmetry and the energy-momentum conservation
of scalar field interacting with a curved background at $D=2$. We
avoid to incorporate the metric determinant into the measure of the
scalar field to explain the conformal anomaly and the consequent
energy-momentum conservation. Contrarily, we split the scalar field in
two other fields, in such a way that just one of them can be
quantized. We show that the same usual geometric quantities of the
anomaly are obtained, which are accompanied by terms containing the
new field of the theory.
\end{abstract}


\maketitle

\section{Introduction}

The interaction between quantum fields and classical gravity has been
intensively studied for a long time \cite{Birrell}. One of the main
motivations is that this procedure is considered to be the first
initial step to understand the full quantum theory including the
gravitatonal field itself. It resembles the old formalism of dealing
with quantum fields interacting with the classical external
electromagnetic background before the advent of the quantum
electrodynamics.

\medskip
However, even though these are aparently similar procedures, the
former is much more involved. For example, the concept of particle is
not well defined. Consequently, the definition of $\vert in\rangle$
and $\vert out\rangle$ states cannot be done in a clear way and the
definition $S$-matrix becomes meaningless. So, instead os particles,
it is the energy-momentum tensor that plays the fundamental role, due
to its local nature and being the source of the curvature in the
general relativity theory. There is an important property that the
energy-momentum tensor has to satisfy, the (covariant) divergenceless,
which corresponds to the conservation of energy and momentum of the
theory. This is a kind of symmetry that cannot be modified by quantum
corrections.

\medskip
On the other hand, we have an interesting symmetry related to massless
theories, called conformal symmetry, which means, in a broader sense,
the absence of scales. This symmetry is manifested in the traceless of
the energy-momentum tensor. Contrarily to previous case, it is not
necessarily kept in the quantum scenario (trace anomaly). This occurs
because the quantum formalism naturally introduces scale parameters in
order to deal with infinities during the regularization procedures.

\medskip
In this paper, we consider a quantum massless scalar field interacting
with a classical curved background. We mention that this theory
exhibits the conformal symmetry, where for $D>2$ it is necessary to
couple the scalar field to the classical curvature (nominimal
coupling) \cite{Birrell}. The particular case of $D=2$, that is the
subject of the present paper, has an interesting feature. The
conformal symmetry is verified without necessity of coupling the
scalar field to the curvature, because there is no conformal
transformation for it. This result may lead to a wrong conclusion that
there is no trace anomaly for scalar fields at $D=2$ because the
absence of conformal transformation for them leads to an invariance of
the corresponding measure in the path integral formalism.

\medskip
The above reasoning and conclusion, which there is no trace anomaly
for $D=2$, cannot be true because it is accompanied by an unpleasant
absence of energy and momentum conservation (after quantum
corrections). A more carefull study of the conformal symmetry shows
that the conformal anomaly does actually exist and the energy and
momentum are actually conserved, as they should be
\cite{Fujikawa,Fujikawa2}. We mention that this problem can be
circumvented by splitting the $\sqrt{-g}$ of the action as $\sqrt{-g}=
(-g)^{-\frac{1}{4}}(-g) ^{-\frac{1}{4}}$ and incorporating each one of
these factors to the scalar field \cite{Fujikawa}. In this way, the
new scalar field acquires a convenient conformal transformation
\cite{Alves}, whose noninvariance of the measure renders the expected
trace anomaly and the energy-momentum conservation.

\medskip
The purpose of the present paper is to display a different alternative
of dealing with this problem. We avoid to incorporate any factor
involving the metric tensor to the scalar field because, since we
intend to work with the path integral formalism, this would be
inconsistent with the initial assumption that the gravitational field
is classical. Our proposal consists in splitting the scalar field in a
product of two fields with different conformal transformations. The
classical conformal symmetry is not modified, but the measure of one
of them is. We show that this leads to a trace anomaly, that has the
same geometrical terms of the usual case, plus other ones related to
the new field. We also show that, eventhough more involved, the energy
momentum conservation is also achieved quantically.

\medskip
Our paper is organized as follows. In Sec. II we brief discuss how
this this problem can be solved by incorporating a factor inovolving
the metric tensor into the scalar field. Eventhough this is a section
review, we follow the lines that we shall used into the next section,
where our formalism is presented. We left Sec. IV for some concluding
remarks.

\section{Trace anomaly and energy-momentum conservation}
\renewcommand{\theequation}{2.\arabic{equation}}
\setcounter{equation}{0}

Let us start from the action

\begin{equation}
S=\frac{1}{2}\int d^2x\,\sqrt{-g}\,g^{\mu\nu}
\partial_\mu\phi\,\partial_\nu\phi
\label{2.1}
\end{equation}

\noindent
The classical energy-momentum tensor is

\begin{eqnarray}
T_{\mu\nu}&=&\frac{2}{\sqrt{-g}}\,\frac{\delta S}{\delta g^{\mu\nu}}
\nonumber\\
&=&\partial_\mu\phi\,\partial_\nu\phi
-\frac{1}{2}\,g_{\mu\nu}\,\partial^\rho\phi\,\partial_\rho\phi
\label{2.2}
\end{eqnarray}

\noindent
which has the following properties

\begin{eqnarray}
&&g^{\mu\nu}T_{\mu\nu}=0
\label{2.3}\\
&&\nabla^\mu T_{\mu\nu}=\partial_\nu\phi\,\Box\,\phi=0
\label{2.4}
\end{eqnarray}

\medskip
\noindent where $\nabla^\mu$ is the covariante derivative and $\Box$
is the Laplace-Beltrami operator, $\Box\phi=\frac{1}{\sqrt{-g}}
\partial_\mu (\sqrt{-g}\partial^\mu\phi)
=g^{\mu\nu}\nabla_\mu\nabla_\nu\phi$.

\medskip
Expression (\ref{2.3}) tell us that the theory exhibits the conformal
symmetry. The conformal transformation for the metric tensor is
$\tilde g_{\mu\nu}=e^{2\alpha}g_{\mu\nu}$ and, consequently,
$\sqrt{-g}\,g^{\mu\nu}$ is conformally invariant. This means that the
conformal symmetry is verified at $D=2$ without any transformation for
the scalar field $\phi$. The meaning of expression (\ref{2.4}) is that
the energy and momentum are conserved. It is opportune to mention that
this expression was obtained by using the equation of motion
for~$\phi$.

\medskip
In the quantum scenario (with a classical background metric) the
energy momentum tensor can be obtained by means of the vacuum
functional as

\begin{equation}
\langle T_{\mu\nu}\rangle
=\frac{2}{\sqrt{-g}}\,\frac{\delta Z}{\delta g^{\mu\nu}}
\label{2.6}
\end{equation}

\noindent
where

\begin{equation}
Z=\,\int[d\phi]\,\exp\,\Bigl(\frac{i}{2}\int d^2x\,
\sqrt{-g}\,g^{\mu\nu}\partial_\mu\phi\partial_\nu\phi\Bigr)
\label{2.5}
\end{equation}

\noindent
Since the scalar field does not change under conformal transformation,
we have that the measure $[d\phi]$ also remains invariant.
Consequently, the semiclassical expression for the energy-momentum
tensor reads

\begin{equation}
\langle T_{\mu\nu}\rangle=
\langle\partial_\mu\phi\,\partial_\nu\phi
-\frac{1}{2}\,g_{\mu\nu}\,\partial^\rho\phi\,\partial_\rho\phi\rangle
\label{2.7}
\end{equation}

\noindent
The curved background is considered to be classical, so we may have

\begin{eqnarray}
&&g^{\mu\nu}\langle T_{\mu\nu}\rangle
=\langle g^{\mu\nu}T_{\mu\nu}\rangle=0
\label{2.8}\\
&&\nabla^\mu\langle T_{\mu\nu}\rangle
=\langle\nabla^\mu T_{\mu\nu}\rangle
=\langle\partial_\nu\,\phi\Box\,\phi\rangle
\label{2.9}
\end{eqnarray}

\noindent
It is not possible to conclude that expression (\ref{2.9}) is zero
because the equation of motion cannot be used in the quantum scenario.
Of course, the results above do not merit confidence because there is
no reason to believe that energy and momentum are not conserved after
quantum effects are taken into account.

\medskip
To circumvent this problem, the action (\ref{2.1}) can be rewriten as

\begin{equation}
S=\frac{1}{2}\int d^2x\,g^{\mu\nu}\partial_\mu\Phi\,\partial_\nu\Phi
\label{2.10}
\end{equation}

\noindent
where

\begin{equation}
\Phi=(-g)^\frac{1}{4}\phi
\label{2.11}
\end{equation}

\noindent
Of course, the classical $T_{\mu\nu}$ is precisely the previous one
given by (\ref{2.2}) (which can be rewritten in terms of $\Phi$) and
the action, $S$ given by (\ref{2.10}), is still conformally invariant.
The conformal transformation for the new scalar field $\Phi$ is

\begin{equation}
\tilde\Phi=e^\alpha\Phi
\label{2.12}
\end{equation}

\noindent
However, for the vacuum functional, the measure $[d\Phi]$ is not
invariant under conformal transformation. In a general way, we have
\cite{Fujikawa3}

\begin{equation}
[d\tilde\Phi]=\exp\Bigl(i\int d^2x\sqrt{-g}\,
\alpha(x)\,A(x)\Bigr)\,[d\Phi]
\label{2.13}
\end{equation}

\noindent
where $A(x)$ is a badly divergent quantity that can be regularized by
means of the zeta function technique leading to \cite{Hawking}

\begin{eqnarray}
A(x)&=&\lim_{s\rightarrow0}\,{\rm tr}\,\zeta(x,s)
\nonumber\\
&=&\frac{[a_1(x)]}{4\pi}
\label{2.14}
\end{eqnarray}

\noindent
The last step of the expression above is restrict to $D=2$, and the
coefficient $a_1(x,x^\prime)$ is related to heat kernel expansion. The
notation $[a_1(x)]$ means $[a_1(x)]= \lim_{x^\prime\rightarrow x}
a_1(x,x^\prime)$. These coefficients can be obtained by means of a
recursion relation that depends on the kind of operator that acts on
the field \cite{DeWitt}. For the present case, $a_1(x)=-\frac{1}{6}
\,R$, where $R$ is the Ricci scalar curvature

\medskip
Now, the corresponding energy-momentum tensor obtained by means of
expression (\ref{2.5}), which we shall denote by $\tilde T_{\mu\nu}$,
is not traceless. The trace $\langle T^\mu\,_\mu\rangle$ can be
directly obtained by

\begin{eqnarray}
\langle\tilde T^\mu\,_\mu\rangle
&=&-\frac{i}{\sqrt{-g}}\,\frac{\delta Z}{\delta\alpha}
\nonumber\\
&=&A(x)
\nonumber\\
&=&-\frac{1}{24\pi}\,R
\label{2.14a}
\end{eqnarray}

\noindent
This result embodies the trace anomaly. Since in two spacetime
dimensions we have the identity $R_{\mu\nu}= \frac{1}{2}g_{\mu\nu}R$,
one may say that the full expression for the energy-momentum tensor
$\langle\tilde T_{\mu\nu}\rangle$ should be

\begin{equation}
\langle\tilde T_{\mu\nu}\rangle
=\langle T_{\mu\nu}\rangle-\frac{1}{48\pi}g_{\mu\nu}R
\label{2.14b}
\end{equation}

\noindent
where $\langle T_{\mu\nu}\rangle$ is the one given by (\ref{2.7}) (it
is indifferent to write it in terms of $\phi$ or $\Phi$). Acting the
covariant derivative in both sides of the expression above, we get

\begin{equation}
\nabla^\mu\langle\tilde T_{\mu\nu}\rangle
=\langle\partial_\nu\phi\,\Box\,\phi\rangle
-\frac{1}{48\pi}\partial_\nu R
\label{2.15}
\end{equation}

\noindent
Expanding the field $\phi$ in terms of eigenfunctions of the operator
$\Box$, one can show that \cite{Fujikawa}

\begin{equation}
\langle\partial_\nu\phi\,\Box\phi\rangle
=\frac{1}{2}\partial_\nu\langle\phi\Box\phi\rangle
\label{2.16}
\end{equation}

\noindent
and the quantity $\langle\phi\,\Box\,\phi\rangle$ can be regularized
and leads to \cite{Fujikawa}

\begin{equation}
\langle\phi\,\Box\,\phi\rangle=\frac{1}{24\pi}\,R
\label{2.18}
\end{equation}

\noindent
So,

\begin{equation}
\nabla^\mu\langle\tilde T_{\mu\nu}\rangle=0
\label{2.19}
\end{equation}

\noindent
as it should be.

\section{Alternative procedure}
\renewcommand{\theequation}{3.\arabic{equation}}
\setcounter{equation}{0}

Now, instead of incorporating the factor $(-g)^{-\frac{1}{4}}$ to the
scalar field, we go in a opposite direction by splitting the field
$\phi$ as

\begin{equation}
\phi=e^\theta\,\varphi
\label{3.1}
\end{equation}

\noindent
where $\theta$ and $\varphi$ are considered to be two independent
quantities with the following conformal tranformations

\begin{eqnarray}
&&\tilde\varphi=e^{-\alpha}\varphi
\label{3.2}\\
&&\tilde\theta=\theta+\alpha
\label{3.3}
\end{eqnarray}

The field $\varphi$ remains quantum, but $\theta$ can be quantum or
not. It is important the field $\theta$ appears in a exponential term
and, consequently, with a conformal transformation like (\ref{3.3}).
This is so because, in the hypothesis that $\theta$ is also quantum,
its corresponding measure, $[d\theta]$, remains unchanged (the
jacobian is trivial). In the developments which follow, we shall
consider $\theta$ classical. At the end, we briefly talk on the
possibility of $\theta$ being quantum.

\medskip
Replacing $\phi$ given by (\ref{3.1}) into the initial expression for
$S$, (\ref{2.1}), we have

\begin{equation}
S=-\frac{1}{2}\int d^2x\,\sqrt{-g}\,
\varphi\,\Bigl[e^{2\theta}
\bigl(\Box -\partial_\mu\theta\partial^\mu\theta)\Bigr]\varphi
\label{3.4}
\end{equation}

We have just done a change of variables and, consequently, there is no
changing into the classical case. But, in the path integral, the
measure $[d\varphi]$ is not invariant under conformal transformation.
Considering $\theta$ classical, se have the vacuum functional

\begin{equation}
Z=\int[d\varphi]\exp\biggl\{-\frac{i}{2}
\int d^2x\sqrt{-g}\varphi\Bigl[e^{2\theta}\bigl(\Box
-\partial_\mu\theta\partial^\mu\theta)\Bigr]\varphi\biggr\}
\label{3.5}
\end{equation}

\noindent
Now, the coefficient $[a_1]$, related to the operator that is acting
on $\varphi$, is

\begin{equation}
[a_1]=-e^{2\theta}\,\Bigl(\frac{1}{6}R
+\partial_\mu\theta\partial^\mu\theta\Bigr)
\label{3.6}
\end{equation}

\noindent
So, the trace anomaly reads

\begin{equation}
\langle T^\mu\,_\mu\rangle=-\frac{e^{2\theta}}{4\pi}\,
\Bigl(\frac{1}{6}R+\partial_\mu\theta\partial^\mu\theta\Bigr)
\label{3.7}
\end{equation}

Since the field $\theta$ is considered to be classical, one may infer
that the expression for the energy momentum tensor
$\langle\tilde T_{\mu\nu}\rangle$ is given by

\begin{equation}
\langle\tilde T_{\mu\nu}\rangle=\langle T_{\mu\nu}\rangle
-\frac{e^{2\theta}}{4\pi}\,
\Bigl(\frac{1}{12}\,g_{\mu\nu}R
+\partial_\mu\theta\partial_\nu\theta\Bigr)
\label{3.8}
\end{equation}

\noindent
where $\langle T_{\mu\nu}\rangle$ is the same one as given by
(\ref{2.7}), with $\phi$ replaced by $e^\theta\varphi$, i.e.

\begin{eqnarray}
&&\langle T_{\mu\nu}\rangle=e^{2\theta}\,
\Bigl[\Bigl(\partial_\mu\theta\partial_\nu\theta
-\frac{1}{2}g_{\mu\nu}\,\partial_\rho\theta\partial^\rho\theta
\Bigr)\langle\varphi^2\rangle
\nonumber\\
&&\phantom{\langle\tilde T_{\mu\nu}\rangle=e^{2\theta}\,}
+\frac{1}{2}\,\partial_\mu\theta\langle\partial_\nu\varphi^2\rangle
+\frac{1}{2}\,\partial_\nu\theta\langle\partial_\mu\varphi^2\rangle
\nonumber\\
&&\phantom{\langle\tilde T_{\mu\nu}\rangle=e^{2\theta}\,}
-\frac{1}{2}\,g_{\mu\nu}\,
\partial_\rho\theta\langle\partial^\rho\varphi^2\rangle
\nonumber\\
&&\phantom{\langle\tilde T_{\mu\nu}\rangle=e^{2\theta}\,}
+\langle\partial_\mu\varphi\partial_\nu\varphi
-\frac{1}{2}g_{\mu\nu}\,
\partial_\rho\varphi\partial^\rho\varphi\rangle\Bigr]
\label{3.9}
\end{eqnarray}

Now, let us verify the consistency with respect the energy and
momentum conservation. First, we consider
$\nabla^\mu\langle T_{\mu\nu}\rangle$. This can be done by directly
acting the operator $\nabla^\mu$ on $\langle\tilde T_{\mu\nu}\rangle$,
given by the expression (\ref{3.9}), or by using (\ref{2.9}), where we
should replace $\phi$ by $e^\theta\varphi$ [notice, however, that this
replacement cannot be done into (\ref{2.16}), because just part of
$\phi$ is quantum]. We thus have

\begin{eqnarray}
&&\nabla^\mu\langle T_{\mu\nu}\rangle=e^{2\theta}\Bigl[
\bigl(\partial_\rho\theta\partial^\rho\theta+\Box\,\theta\bigr)\,
\partial_\nu\theta\langle\varphi^2\rangle
\nonumber\\
&&\phantom{\nabla^\mu\langle\tilde T_{\mu\nu}\rangle
=e^{2\theta}\Bigl[}
+\partial_\nu\theta\partial^\rho\theta\,
\langle\partial_\rho\varphi^2\rangle
+\partial_\nu\theta\langle\varphi\,\Box\,\varphi\rangle
\nonumber\\
&&\phantom{\nabla^\mu\langle\tilde T_{\mu\nu}\rangle
=e^{2\theta}\Bigl[}
+\frac{1}{2}\bigl(\partial_\rho\theta\partial^\rho\theta
+\Box\,\theta\bigr)\,\langle\partial_\nu\varphi^2\rangle
\nonumber\\
&&\phantom{\nabla^\mu\langle\tilde T_{\mu\nu}\rangle
=e^{2\theta}\Bigl[}
+2\partial^\rho\theta
\langle\partial_\rho\varphi\partial_\nu\varphi\rangle
+\langle\partial_\nu\varphi\,\Box\,\varphi\rangle\Bigr]
\label{3.10}
\end{eqnarray}

\noindent
For the second term of (\ref{3.8}) we have

\begin{eqnarray}
&&-\frac{1}{4\pi}\,\nabla^\mu\,
\Bigl[e^{2\theta}\,\Bigl(\frac{1}{12}\,g_{\mu\nu}\,R
+\partial_\mu\theta\partial_\nu\theta\Bigr)\Bigr]
\nonumber\\
&&\phantom{-\frac{1}{4\pi}\,\nabla^\mu\,}
=-\frac{1}{4\pi}\,e^{2\theta}\,
\Bigl(\frac{1}{6}\,R\,\partial_\nu\theta
+\frac{1}{12}\,\partial_\nu R
\nonumber\\
&&\phantom{-\frac{1}{4\pi}\,\nabla^\mu\,=}
+2\partial_\nu\theta\partial_\rho\theta\partial^\rho\theta
\nonumber\\
&&\phantom{-\frac{1}{4\pi}\,\nabla^\mu\,=}
+\Box\,\theta\partial_\nu\theta
+\partial^\rho\theta\partial_\rho\partial_\nu\theta\Bigr)
\label{3.11}
\end{eqnarray}

\noindent
Now, since just $\varphi$ is quantum, we can write

\begin{equation}
\langle\partial_\nu\varphi\,\Box\,\varphi\rangle
=\frac{1}{2}\partial_\nu\langle\varphi\,\Box\,\varphi\rangle
\label{3.12}
\end{equation}

\noindent
Considering the action involving $\varphi$ and with the term
$e^{2\theta}$ factorized, as well as the renormalization factor we are
using into (\ref{2.14}), we can obtain the regularized quantities

\begin{eqnarray}
&&\langle\varphi\,\Box\,\varphi\rangle
=\frac{1}{4\pi}\Bigl(\frac{1}{6}\,R
+\partial_\rho\theta\partial^\rho\theta\Bigr)
\label{3.13}\\
&&\langle\varphi^2\rangle=\frac{1}{4\pi}
\label{3.14}
\end{eqnarray}

\noindent
Finally, since $\varphi$ is a scalar quantity and
$\langle\partial_\nu\varphi^2\rangle$ is an average involving all
directions aleatory, and the same occurs with
$\langle\partial_\rho\varphi\partial_\nu\varphi\rangle$, we have that
these quantities are null. Replacing all these results into
(\ref{3.10}) and (\ref{3.11}), we get

\begin{equation}
\nabla^\mu\langle\tilde T_{\mu\nu}\rangle=0
\label{3.15}
\end{equation}

\noindent
which express the consistency with the energy-momentum conservation.

\medskip
In the case that $\theta$ is also quantum, the measure $[d\theta]$
does not change under conformal transformation, but the problem is
much more involved and difficult to solve. Just
formally, one may write that the trace anomaly reads

\begin{equation}
\langle\tilde T^\mu\,_\mu\rangle=-\frac{1}{4\pi}\,
\langle e^{2\theta}\Bigl(\frac{1}{6}R
+\partial_\mu\theta\partial^\mu\theta\Bigr)\rangle
\label{3.16}
\end{equation}

\noindent
From which one cannot either infer an expression similar as the second
term of (\ref{3.8}) or try to regularize it because the bad
divergencies occuring in the exponential $e^{2\theta}$.

\section{Conclusion}

In this paper we have study the problem of conformal anomaly and the
energy-momentum conservation for a quantum scalar field interacting
with a classical curved background in the spacetime dimension D=2. We
have considered the scalar field as split in two other scalar fields,
and kept just of them quantum. This procedure is in opposite
direction to what is done in literature, where a factor containing the
metric determinant is absorbed by the scalar field, leading to a new
field with a convenient conformal transformation. We have shown that
our procedure is consistent with the geometric terms of the usual
treatment of the conformal anomaly and also consistent with the
expected result that energy and momentum should be conserved after
quantum effects are taken into account.

\begin{acknowledgments}
This work is supported in part by Conselho Nacional de Desenvolvimento
Cient\'{\i}fico e Tecnol\'ogico - CNPq  and Funda\c{c}\~ao
Universit\'aria Jos\'e Bonif\'acio - FUJB (Brazilian Research
Agencies) One of us, J.B.-N. has also the support of PRONEX
66.2002/1998-9.
\end{acknowledgments}

\end{document}